\documentclass[a4paper,pra,twocolumn, superscriptaddress]{quantumarticle}
\pdfoutput=1
\usepackage{hyperref}
\usepackage{amssymb,amsmath}
\usepackage{mathtools}
\usepackage{ifxetex,ifluatex}
\usepackage{graphicx}
\usepackage{float}
\hypersetup{breaklinks=true, pdfborder={0 0 0}}

\begin{document}

\title{Quantum computed moments correction to variational estimates}

\author{Harish J. Vallury}
\author {Michael A. Jones}
\affiliation{School of Physics, University of Melbourne, Parkville 3010, AUSTRALIA}

\author{Charles D. Hill}
\affiliation{School of Physics, University of Melbourne, Parkville 3010, AUSTRALIA}
\affiliation{School of Mathematics and Statistics, University of Melbourne, Parkville 3010, AUSTRALIA}

\author{Lloyd C. L. Hollenberg}
\email{lloydch@unimelb.edu.au}
\affiliation{School of Physics, University of Melbourne, Parkville 3010, AUSTRALIA}

\begin{abstract}
{The variational principle of quantum mechanics is the backbone of hybrid quantum computing for a range of applications. However, as the problem size grows, quantum logic errors and the effect of barren plateaus overwhelm the quality of the results. There is now a clear focus on strategies that require fewer quantum circuit steps and are robust to device errors. Here we present an approach in which problem complexity is transferred to dynamic quantities computed on the quantum processor -- Hamiltonian moments, $\langle H^n\rangle$. From these quantum computed moments, an estimate of the ground-state energy can be obtained using the ``infimum'' theorem from Lanczos cumulant expansions which manifestly corrects the associated variational calculation. With higher order effects in Hilbert space generated via the moments, the burden on the trial-state quantum circuit depth is eased. The method is introduced and demonstrated on 2D quantum magnetism models on lattices up to $5\times 5$ (25 qubits) implemented on IBM Quantum superconducting qubit devices. Moments were quantum computed to fourth order with respect to a parameterised antiferromagnetic trial-state. A comprehensive comparison with benchmark variational calculations was performed, including over an ensemble of random coupling instances. The results showed that the infimum estimate consistently outperformed the benchmark variational approach for the same trial-state. These initial investigations suggest that the quantum computed moments approach has a high degree of stability against trial-state variation, quantum gate errors and shot noise, all of which bodes well for further investigation and applications of the approach.}

\end{abstract}

\maketitle

\section{Introduction}
Quantum computers represent a new paradigm for computing that is witnessing rapid advances in both hardware and software. Fully programmable devices are emerging, and evidence of information processing at a scale that competes with supercomputers for certain sampling problems has been reported in a quantum device comprising 53 qubits \cite{Arute-2019}. The major challenge of the field is to demonstrate ``quantum advantage'' for real-world problems. There are approaches to a range of potential application areas, including bioinformatics \cite{LH-2000}, chemistry \cite{Aspuru-Guzik-2005,Kandala-2017}, optimisation \cite{Farhi-2014}, finance \cite{Rebentrost-2018, Woerner-2019} and machine learning \cite{Biamonte-2017}, to name a few. However, in the short to medium term, quantum computer (QC) technology will be constrained to the so-called noisy intermediate scale quantum (NISQ) regime \cite{Preskill-2018} -- where the performance of QC devices will inevitably be dominated by the level of logic precision inherent in the hardware. 

\begin{figure*}[htb]
\includegraphics[width=18cm]{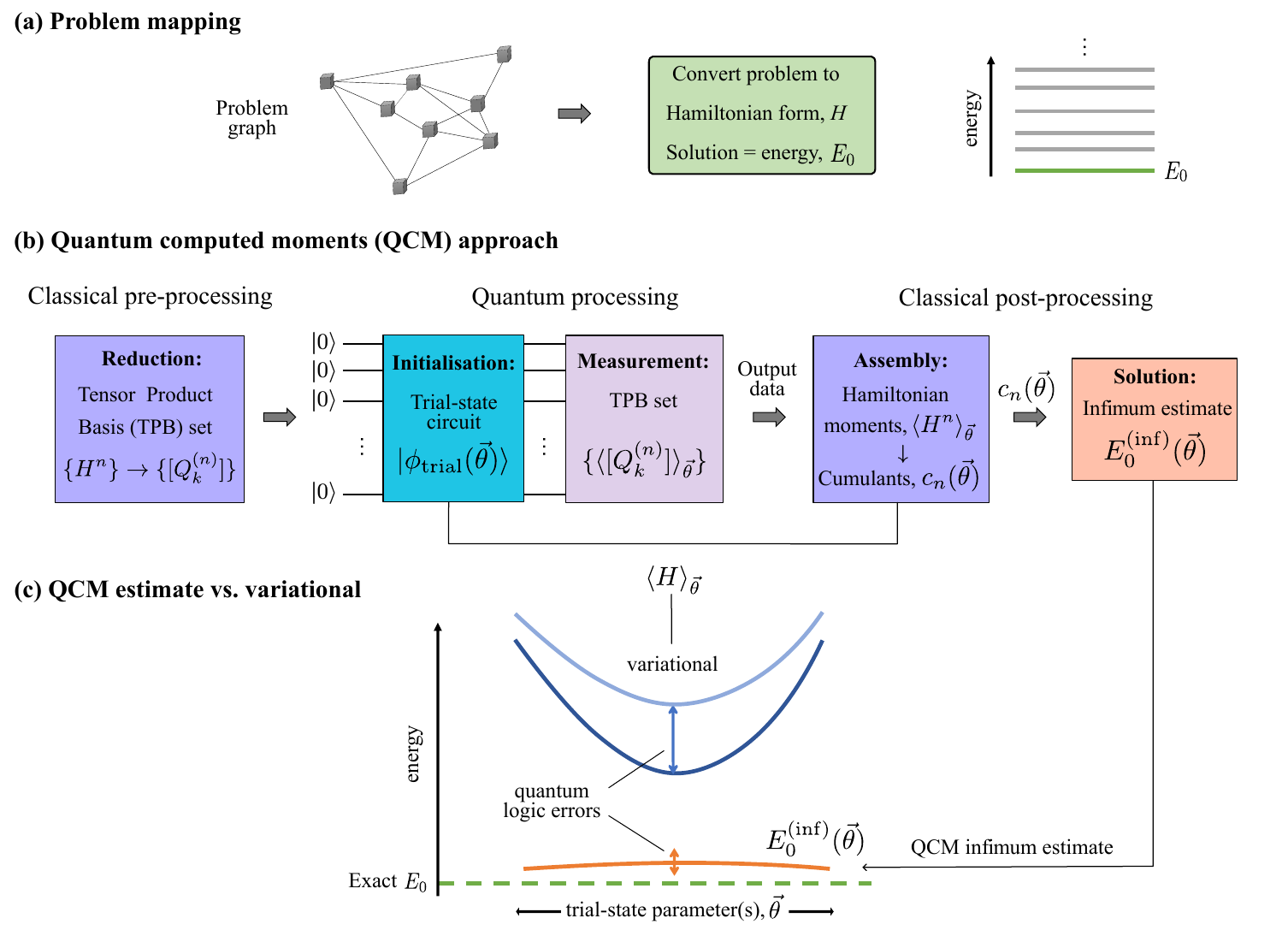}
\caption{Overview of the quantum computed moments (QCM) approach. (a) Problem cast in Hamiltonian form, $H$, with solution energy $E_0$. (b) Pre-processing: reduction of the exponentiated form $H^n$ into tensor product basis (TPB) sets of Pauli strings $\{Q^{(n)}_k\}$, to order $n_{\rm max}$. Quantum processing: with respect to a trial-state $|\phi_{\rm trial}(\vec{\theta})\rangle$ the expectation values $\langle Q^{(n)}_k\rangle$ of all terms in the TPB set are measured. Post-processing: from the measurements of $\langle Q^{(n)}_k\rangle$, Hamiltonian moments, $\langle H^n\rangle \equiv \langle \phi_{\rm trial}(\vec{\theta})|H^n|\phi_{\rm trial}(\vec{\theta})\rangle$ are assembled, transformed to cumulants $c_n(\vec{\theta})$, and the infimum estimate $E_0^{(\rm inf)}(\vec{\theta})$ computed. (c) The infimum estimate $E_0^{(\rm inf)}(\vec{\theta})$ of the ground-state energy solution, $E_0$, provides a correction to the variational result ${\rm min}\langle H\rangle_{\vec{\theta}}$.}
\label{Fig1}
\end{figure*}

In NISQ devices, the quest for quantum advantage is challenged by errors in logic and read-out which place severe restrictions on the number of time-steps, or ``depth'', of any given quantum circuit before the results are scrambled. Hybrid quantum algorithms such as the Variational Quantum Eigensolver (VQE) \cite{Peruzzo-2014} or the Quantum Approximate Optimisation Algorithm (QAOA) \cite{Farhi-2014} adapt variational-style hybrid approaches to the problem cast in Hamiltonian form. However, the application to real-world problems generally requires relatively deep quantum circuits leading to barren plateau effects, exacerbated by the accumulation of quantum logic errors \cite{McClean-2018,Wang-2020}. In NISQ devices, quantum circuit depth is perhaps the most precious quantum resource. 

In this work we introduce an alternative method for computing the lowest energy of a problem Hamiltonian system, $H$, based on minimising circuit depth by transferring complexity to the computation of moments of the Hamiltonian, $\langle H^n\rangle$, with respect to a given trial-state (Figure \ref{Fig1}). Such quantities are central to certain non-perturbative approximation schemes in many-body theory, but are generally difficult to compute classically as the problem scales. The key in our approach is to reserve the quantum resource for the direct computation of the Hamiltonian moments, which  are then used to determine  ground-state energy estimates that correct the variational result (first moment, $\langle H\rangle$). In our initial implementation of the quantum computed moments (QCM) method, the results obtained appear highly robust to device noise and provide a significant correction to the energy of the benchmark variational result. 

This paper is organised as follows: Section \ref{sec:Variational} describes the problem in Hamiltonian form and the variational approach in the context of hybrid quantum computing. In Section \ref{sec:Lanczos} we review the background of the QCM approach: the connection of Hamiltonian moments to the Lanczos expansion approach and the infimum theorem that provides energy estimates to a given moment order. In Section \ref{sec:Problem} we define the problem Hamiltonian for the demonstration -- quantum spin systems on 2D lattices -- and provide details on the tensor product basis set construction and scaling. Results obtained by computing moments on IBM Quantum devices and associated simulations are provided in Section \ref{sec:Results}, and conclusions drawn in Section \ref{sec:Conclusion}.

\vspace{5mm}

\section{Hamiltonian form and the variational limit} \label{sec:Variational}

We consider problems which, when converted to a quantum context, can be reduced to finding the ground-state energy of an equivalent problem Hamiltonian $H$ over strings of qubit operators $\{I, X, Y, Z\}$ -- this general class of problems encompasses a number of applications. We write the problem Hamiltonian over $q$ qubits as:
\begin{equation}
H = \sum_{i} w_i [P_i].
\end{equation}
The sum is over Pauli strings $[P_i] \equiv \sigma_1(i)\sigma_2(i)...\sigma_{q}(i)$, where for the $m$th qubit we have $\sigma_m(i) = (I, X, Y,  {\rm or} \,Z)$ defined by the problem, and  each term has a weight $w_i$. The solution is the lowest energy state of $H$, which we denote $E_0$. This is, in general, a difficult task given the Hilbert space dimension grows as $2^{q}$. 

Approaches to solve such problems based on the well-known variational principle in quantum mechanics have gained widespread appeal in the NISQ era of quantum computing. The variational quantum eigensolver (VQE) \cite{Peruzzo-2014} is a hybrid approach to finding an approximation to $E_0$ on a quantum computer. VQE begins by setting up a quantum circuit to create the parameterised trial-state $|\phi_{\rm trial}(\vec{\theta})\rangle$ over a set of parameters $\vec{\theta}$. The expectation values $\langle [P_i]_{\vec{\theta}} \rangle$, in the state $|\phi_{\rm trial}(\vec{\theta})\rangle$ are estimated term by term via repeated initialisation and measurement of the QC, and summed to produce $\langle H\rangle_{\vec{\theta}}$. This procedure is incorporated into a classical loop minimising the result with respect to the trial-state parameters to produce the lowest upper bound $\langle H\rangle_{\vec{\theta}}$ on $E_0$. In VQE, the state $|\phi_{\rm trial}(\vec{\theta})\rangle$ is implemented as a sequence of $p$ sub-circuits involving mixing and entangling operations governed by parameters $\vec{\theta}(m)$ in the $m$th sub-circuit block. As the number of circuit blocks increases, the total circuit depth grows, as does the set of parameters $\{\vec{\theta}(1),\vec{\theta}(2),...\vec{\theta}(p)\}$, the trial-state becomes more and more complicated in order to better approximate the ground-state of the problem. Working against this, errors in a NISQ device accumulate in the output, thereby restricting the maximum depth of the trial-state circuit and limiting convergence of the variational procedure \cite{McClean-2018,Wang-2020}. This is generally a critical barrier to overcome in practical applications. 

While the variational procedure is an obvious and time-proven place to start, it is clear that the quantum resource must be used as efficiently as possible. We focus on valuing quantum circuit depth by exploiting dynamics in the Hamiltonian as the definitive generator of the ground-state estimate to ease the complexity of the trial-state. The quantum computed moments approach produces a correction to the variational result, effectively trading quantum circuit depth for an increased number of measurements and classical post-processing. 
\section{Quantum computed moments (QCM) approach} \label{sec:Lanczos}

The focus on using a quantum computer to directly compute moments of the Hamiltonian $\langle H^n\rangle$ with respect to a trial-state is inspired by a cumulant expansion of the Lanczos method \cite{LH-1993}. Uncovered for extensive systems in the context of lattice gauge field theory, the cumulant expansion of the Lanczos tri-diagonalised form allows for the ground-state of the diagonalised system to be obtained via an ``infimum'' theorem \cite{LH-1996}. Typically, the computation of $\langle H^n\rangle$ scales poorly with classical computing resources for non-homogenous $H$ -- the utility of the infimum approach has been limited to homogeneous systems. However, the emergence of quantum computers sheds new light on the overall approach and its possibilities. 

We begin the description with the well-known Lanczos recursion method  \cite{Lanczos-1950} for estimating the lowest energy eigenvalue(s) of a system. It has recently been considered in the quantum computing context with the introduction of ``imaginary-time'' evolution algorithms \cite{McArdle-2019,Motta-2020, Yeter-Aydeniz-2020} and in  error mitigation \cite{Suchsland-2020}. The transformation of the Hamiltonian into tri-diagonal form with respect to some initial trial-state $|v_1\rangle$, proceeds according to the Lanczos recursion as \cite{Lanczos-1950}:
\begin{equation}
|v_{i}\rangle = {1\over \beta_{i-1}}\biggl[(H - \alpha_{i-1})| v_{i-1}\rangle - \beta_{i-2}| v_{i-2}\rangle\biggr].
\end{equation}
The matrix elements of the Hamiltonian in the Lanczos-basis are given by $\alpha_i = \langle v_i|H|v_i\rangle$ and $\beta_i = 
\langle v_{i+1}|H|v_i\rangle$. In classical applications, the eigenvalues of the truncated tri-diagonal Hamiltonian matrix in the Lanczos-basis $|v_i\rangle$ are computed numerically. Although the approximates converge relatively rapidly to the low lying states of the original Hamiltonian, the classical computation of the $\alpha_i$ and $\beta_i$ are generally limited by the matrix dimension of the problem. For the quantum computing context we consider the Lanczos recursion in the formalism of Hamiltonian moments $\langle H^n\rangle \equiv \langle v_1|H^n|v_1\rangle$, with respect to an appropriate initial state, $|v_1\rangle$ (e.g. in the ground state sector of the Hilbert space). Some time ago, it was found that there exists a general expansion of the Lanczos matrix elements with respect to the Hamiltonian moments \cite{LH-1993,LH-1996}:
\begin{eqnarray}
& &\alpha(z) =
c_1 + z\,\left[{c_{3}\over c_{2}}\right]
 + z^2\,\left[{3 c_3^3 - 4 c_2 c_3
c_4 + c_2^2 c_5 \over 4 c_2^4}\right] + ...,\nonumber
\\\nonumber\\
& &\beta^2(z) = z\,c_{2} + z^2\left[{c_{2}c_{4} - c_{3}^{2}\over 2 c_{2}^{2}}\right] + ...
\end{eqnarray}
where $z$ is a continuous positive parameter related to the recursion index \cite{LH-1996,LH-1996-2} ($z\equiv i/V$ for extensive systems of volume $V$), and the quantities $c_n$ are the cumulants derived from the moments $\langle H^n\rangle$:  
\begin{equation}
c_n = \langle H^n\rangle - \sum_{p=0}^{n-2}{n-1\choose p}c_{p+1} \langle H^{n-1-p}\rangle .
\end{equation}
The explicit general expressions for the Lanczos matrix elements $\alpha(z)$ and $\beta(z)$ allow powerful results from orthogonal polynomial theory to be brought into play   \cite{VanDoorn-1987,Ismail-1992}. As a result, at all orders in the $z$-expansion for the Lanczos matrix elements, the ground-state energy of the Hamiltonian system can be expressed directly via an infimum theorem \cite{LH-1996}:
\begin{equation}
E_0 =\inf_{z > 0}\,\left[\alpha(z) - 2\,\beta(z)\right].
\end{equation}
This rather innocuous looking relationship actually diagonalises the tri-diagonal Lanczos system in its expanded form, allowing one to determine approximates, $E_0^{(\rm inf)}$, to the solution $E_0$ by truncating the $z$-expansion to some maximum moment/cumulant order $n_{\rm max}$. At first order in $z$, a general ``infimum'' approximate for the ground state energy involving cumulants to order $n_{\rm max} = 4$ was derived \cite{LH-1994}: 
\begin{eqnarray}
E_0^{(\rm inf)} = c_1 - {c_2^2\over c_3^2 - c_2 c_4} \left[\sqrt{3 c_3^2 - 2 c_2 c_4} - c_3\right].
\end{eqnarray}
The expression for $E_0^{({\rm inf})}$ represents the exact diagonalisation of the system with respect to a finite set of cumulants (up to $c_4$ in this case), with the second term in Eqn (6) providing the corresponding correction to the energy $c_1 = \langle H\rangle$ of the trial-state $|\phi_{\rm trial}\rangle \equiv |v_1\rangle$. It is worthwhile to note that $E_0^{({\rm inf})}$ is no longer a strict upper bound, and the relationship of the cumulant-based correction to the variational result $c_1 = \langle H\rangle$ is governed by the overlap of the trial-state with the true ground state. In principle, one can determine higher order infimum estimates \cite{Witte-1997}, however, here we will show the analytic $E_0^{(\rm inf)}$ expression for $n_{\rm max} = 4$ already provides a convenient and powerful correction to the variational calculation represented explicitly by $c_1$. Higher order contributions in Hilbert space driven by the Hamiltonian of the system are neatly encapsulated in the cumulants, and since the correction to the variational estimate of $E_0$ obtained from the infimum theorem corresponds to the diagonalised Hamiltonian in the tridiagonal Lanczos basis, derived approximates sum the associated (truncated) dynamical effects to all orders. The Lanczos expansion and associated infimum theorem has been applied to a number of homogenous many-body systems, from quantum magnetism \cite{Witte-1997-2} to lattice gauge theory \cite{McIntosh-2002}, however, the computation of moments for non-homogenous systems  generally scale poorly on classical resources as the system size grows. For systems of interest we seek to directly compute these quantities on a quantum processor. 

We thus arrive at the quantum computed moments (QCM) approach for estimating the lowest energy of the problem Hamiltonian:
\begin{enumerate}
\item Problem mapping: $H$ $\leftrightarrow$ solution $E_0$
\item Reduction: $H^n\rightarrow \{[Q_k^{(n)}]\}$ TPB set to order $n_{\rm max}$
\item Design/prepare trial-state: $|\phi_{\rm trial}(\vec{\theta})\rangle$ 
\item Quantum compute $\langle [Q_k^{(n)}]\rangle_{\vec{\theta}}$ in $|\phi_{\rm trial}(\vec{\theta})\rangle$
\item Assemble moments $\langle H^n\rangle_{\vec{\theta}}$ 
\item Assemble cumulants $c_n(\vec{\theta})$ (at min $c_1(\vec{\theta})$)
\item Obtain infimum approximate $E_0^{(\rm inf)}(\vec{\theta})$ to $E_0$.
\end{enumerate}

\begin{figure*}[htb]
\includegraphics[width=17cm]{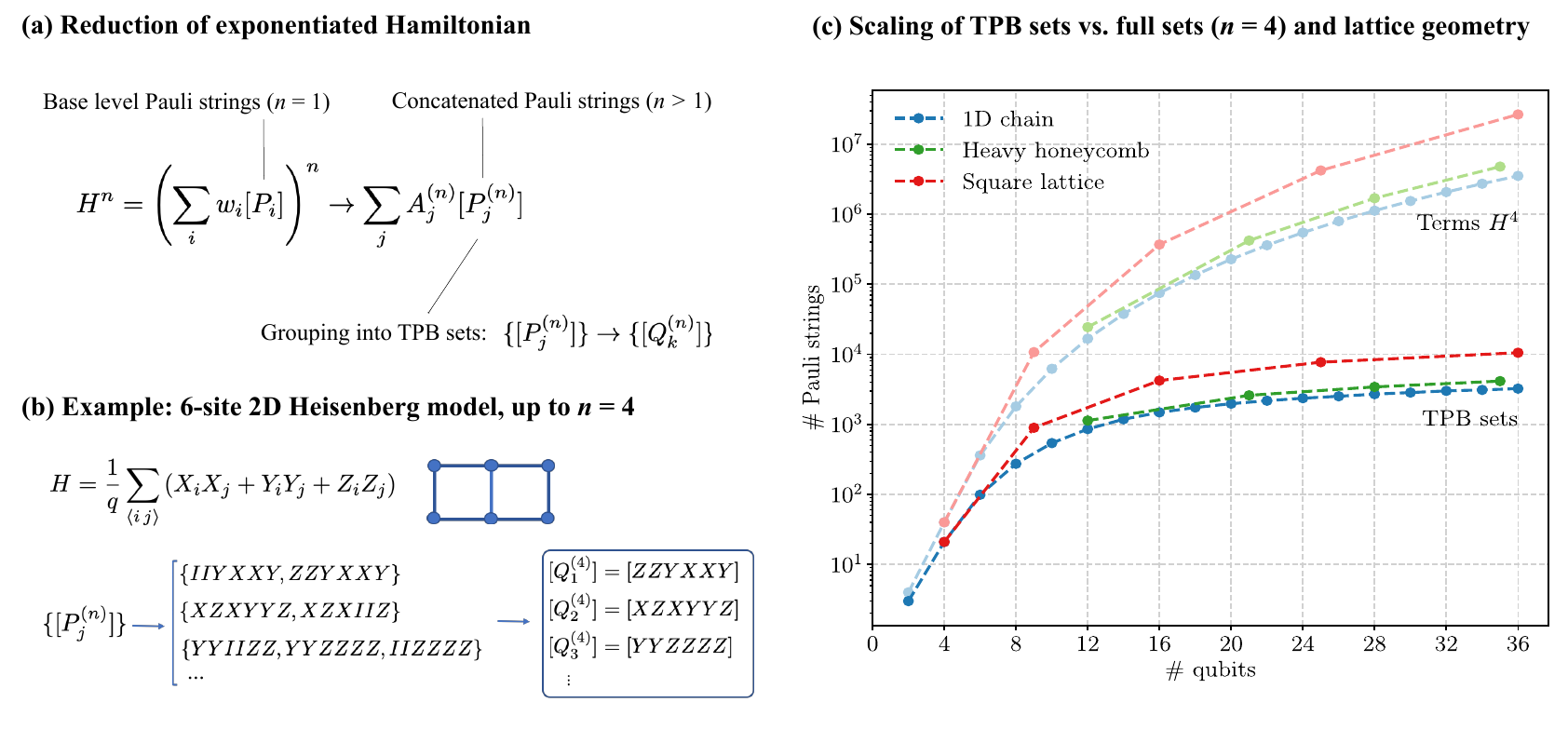}
\caption{Overview of Hamiltonian exponentiation. (a) Grouping of concatenated Pauli strings $[P_j^{(n)}]$ in $H^n$ into Tensor Product Basis (TPB) sets $[Q_k^{(n)}]$. (b) Example: the first few TPB sets for the $q=6$ 2D Heisenberg model. (c) Operator term counts for $H^4$ before and after TPB grouping for the quadratic model defined on a 1D chain, heavy-honeycomb and square lattice.}
\label{Fig2}
\end{figure*}

Before moving to implementation, we make a few general remarks. The QCM approach places more emphasis on the number of circuit runs on the quantum computer and associated classical computation to achieve a better energy estimate rather than increasing the circuit depth in the variational approach. The actual complexity of the quantum information processing task involved in computing moments relative to classical resources is dependent on the details of the trial-state, as we note further on. At first sight, the procedure for exponentiating the Hamiltonian appears to scale badly, however, the number of terms can be tightly controlled by creating Tensor Product Basis (TPB) sets. Another issue that we focus on in the implementation is how shot noise and device errors flow through the arithmetical operations from TPB set measurements, to moments, to cumulants. Although we have articulated the approach in the context of obtaining infimum estimates for the ground-state energy, we note the quantum computed moments may be used beyond this context (see Conclusion). The approach here is quite distinct to recently proposed Lanczos and power methods \cite{McArdle-2019,Motta-2020, Yeter-Aydeniz-2020,Suchsland-2020,Seki-2020} and approximates in the connected moment expansion form of the $t$-expansion   \cite{Horn-1984,Cioslowski-1987,Kowalski-2020}. The infimum estimate corresponds to the diagonalisation of a truncated cumulant expansion of the Lanczos tri-diagonal basis, as opposed to the usual approach of computing and diagonalising the truncated Lanczos basis itself, e.g. the basis of the scheme proposed in \cite{Suchsland-2020}. 

\section{Hamiltonian problem, operator reduction and scaling} \label{sec:Problem}

To show how the algorithm performs in practice, we will consider a non-trivial example system from quantum magnetism. The quadratic Hamiltonian (density) for $q$ qubits is given by:
\begin{equation}
H = {1\over q} \sum_{\langle i\,j\rangle} \left(J_{ij}^{(x)} X_i X_j + J_{ij}^{(y)} Y_i Y_j + J_{ij}^{(z)} Z_i Z_j\right),
\end{equation}
where the sum is over a problem graph defined by the vertices (qubits) $ {i =1...q}$, edges connecting qubits $\{\langle i\,j\rangle\}$, and couplings $J_{ij}^{(s)}$ along each edge ($s = x, y, z$). Here we consider nearest-neighbour 2D lattices with free-boundary conditions. The uniform coupling case $J_{ij}^{(x)} = J_{ij}^{(y)} = J_{ij}^{(z)}$ is the well known 2D Heisenberg model, for which the problem of computing the exact ground state has been extensively studied. While the bipartite graph case can be mapped to a stoquastic Hamiltonian \cite{Bravyi-2006}, the QCM technique can be used in general for harder Hamiltonian models, in particular Heisenberg models which are QMA complete \cite{Cubitt-2016}.

\begin{figure*}[htb]
\includegraphics[width=17cm]{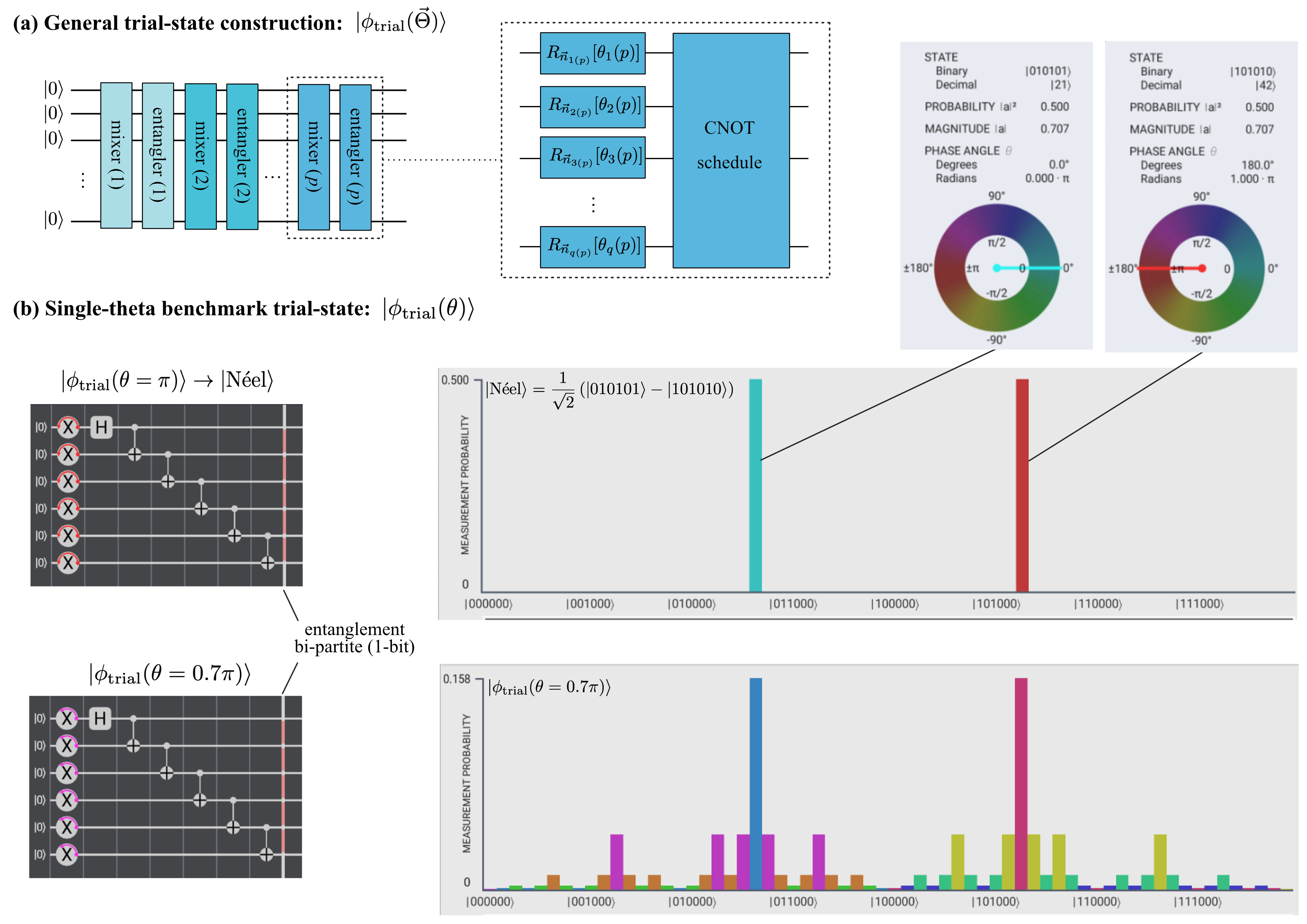}
\centering
\caption{Trial-state circuit construction. (a) General multi-parameter $p$-block trial-state construction for hybrid variational approaches. (b) Quantum circuits for initialising the trial-state $|\phi_{\rm trial}(\theta)\rangle$ over a 1D qubit array for the 2D $2\times 3$ problem: $\theta = \pi$ (N\'eel state) and $\theta = 0.7 \pi$. Vertical bar at right indicates the level of bi-partite entanglement-entropy in the (ideal) trial-state at 1-bit. Probability distributions are shown together with the overall phase of the individual state amplitudes (visualisation using the quantum user interface (QUI) system \cite{quispace-2018})}. 
\label{Fig3}
\end{figure*}

We first detail the Hamiltonian exponentiation and the scaling of the effective number of Pauli strings required for measurement. Initially, one concatenates and compresses products of Pauli strings at each level of $H^n$:
\begin{equation}
H^n = \left(\sum_{i} w_i [P_i]\right)^n \xRightarrow{\text{concatenation}} \sum_{j} A_j^{(n)} [P_j^{(n)}],
\end{equation}
where the $[P_j^{(n)}]$ are $q$-length Pauli strings resulting from the product reductions, and $A_j^{(n)}$ are the resulting weights. Naive counting suggests the number of Pauli strings in the expressions corresponding to powers of the Hamiltonian increases exponentially with $n$. However, by exploiting the properties of the Pauli matrices and their commutation relations, the number of strings required for measurement can be drastically reduced by finding tensor product basis (TPB) sets $[Q_k]$ of Pauli strings that mutually qubit-wise commute (QWC) \cite{Kandala-2017}. Thus, we rewrite the operator $H^n$ in terms of the TPB sets as: 
\begin{equation}
H^n = \sum_{j} A_j^{(n)} [P_j^{(n)}]: \{[P_j^{(n)}]\} \xRightarrow{\text{TPB grouping}} \{[Q_k^{(n)}]\},
\end{equation}
where products of Hamiltonian-level strings $[P_j^{(n)}]$ in $H^n$ are grouped to form TPB sets of QWC Pauli strings, $[Q_k^{(n)}]$, which are labeled by the string $[Q_k^{(n)}]$ itself (see Figure \ref{Fig2}(a) and (b)). Measurement need only be carried out over the $[Q_k^{(n)}]$, reducing the overall measurement burden accordingly. 

\begin{figure*}[htb]
\includegraphics[width=17cm]{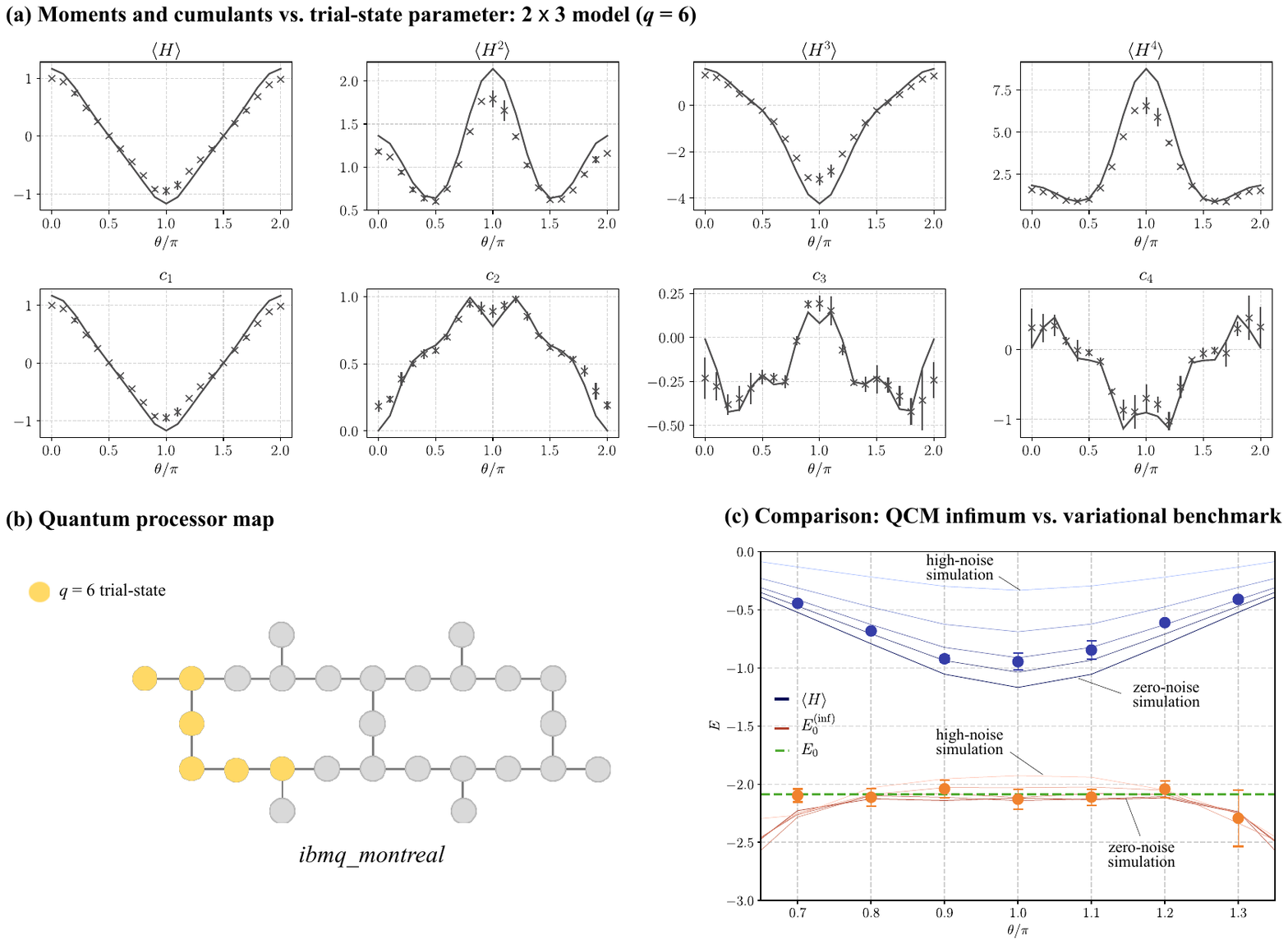}
\centering
\caption{Moments-based quantum computing algorithm applied to the $q=6$ 2D Heisenberg model (uniform couplings) with respect to the single $\theta$ trial-state (Fig \ref{Fig3}). (a) Quantum computed Hamiltonian moments and cumulants, respectively, assembled from the measurement of the expectation values of the TPB Pauli strings $\langle [Q_k^{(n)}]\rangle$. Data points correspond to calculations on the IBM Quantum processor {\it ibmq\_montreal} with statistical error bars corresponding to $5\times 1024$ shots per expectation value. Solid grey lines  are simulations carried out using the Qiskit QASM simulator at zero noise. (b) Quantum processor device map for {\it ibmq\_montreal} showing the qubits used in the computation. (c) QCM infimum estimate obtained from the device runs (orange data points), plotted as a function of the trial-state parameter $\theta$, directly compared with the benchmark variational results derived from $\langle H\rangle_{\theta}$ (blue data points). The exact value is shown as a green dashed line. Orange and blue solid lines correspond to simulations at different noise levels, from zero to a high-noise scenario in steps of 2$\times$ the default device error model.}
\label{Fig4}
\end{figure*}

Finding the optimal TPB sets $[Q_k{(n)}]$ can be mapped to a minimum clique cover problem for the equivalent graph and solved heuristically \cite{Verteleskyi-2020}. Here we employ an identity-operator sorting algorithm. Pauli strings are sorted into a growing collection of TPB sets $[Q_k{(n)}]$, starting with the Pauli strings which have the lowest number of identities. Each $[Q_k{(n)}]$ is created when a Pauli string does not QWC with any other $[Q_k{(n)}]$ in the collection. When a Pauli string $P$ does QWC with an existing $[Q_k{(n)}]$, it is assigned to the corresponding TPB set, and identities in the string $[Q_k{(n)}]$ change to one of ${X,Y,Z}$ accordingly to ensure that any new Pauli string added to the set is QWC with $P$. The final TPB sets $[Q_k^{(n)}]$ of Pauli strings to be measured depends on the underlying problem graph $\{\langle i\,j\rangle\}$. To show this, we perform the reduction process for three types of graphs -- linear, heavy-honeycomb, and square lattice -- for systems defined on up to 36 qubits. In Figure \ref{Fig2}(c) we plot the growth with the number of qubits, $q$, for naive term counting. The dramatic effect of the TPB grouping process on the scaling is evident -- for a given $q$, the number of Pauli strings to be measured drops by several orders of magnitude with sub-linear scaling in $q$. The qubit-wise measurements of each of the concatenated strings $[P_j^{(n)}]$ in the TPB set $[Q_k^{(n)}]$ are reconstructed from the qubit-wise measurements of the string $[Q_k^{(n)}]$. Repeated measurements are summed to produce expectation values of the Pauli strings $[P_j^{(n)}]$ from which the moments $\langle H^n\rangle$ are assembled. 

\section{Results} \label{sec:Results}

As a first application of the method we consider the case of the 2D Heisenberg model defined on a $2\times 3$ lattice with uniform coupling $J_{ij}^{(s)} =1$. The first few TPB sets produced by the grouping algorithm at order $H^4$ are shown in Figure \ref{Fig2}(b). Following a simplified version of the VQE construction Fig \ref{Fig3}(a), we define a trial-state in single parameter form, $|\phi_{\rm trial}(\theta)\rangle$, as shown in Fig \ref{Fig3}(b). This choice of trial-state includes the antiferromagnetic N\'eel state at $\theta = \pi$. Away from $\theta = \pi$ the full set of $2^q$ states are engaged (e.g. $\theta=0.7\pi$ shown). While the model itself is defined on a 2D lattice, we meet the challenge of restricted qubit array connectivity by defining the trial-state over a 1D array of qubits.

Results shown in Fig \ref{Fig4} correspond to the QCM algorithm run on the IBM Quantum processor {\it ibmq\_montreal} -- the qubits used are indicated on the device map. Comparison simulations were run on the Qiskit QASM simulator \cite{qiskit}. The default error model used incorporates depolarizing error and readout error to match those of the available devices. The single-qubit depolarizing error rate is set to 0.001, and the two-qubit depolarizing error rate is set to 0.01. For readout errors we set Prob(0$|$1) = 0.03 and Prob(1$|$0) = 0.015. We have plotted, as a function of the trial-state parameter $\theta$, the moments $\langle H^n\rangle$, and associated cumulants $c_n$, assembled from the QC measurements of the TPB sets $\{[Q_k^{(n)}]\}$ up to $n_{\rm max}=4$. Quantum calculations were carried out using $5\times1024$ shots per expectation value. Note: these are raw results with no attempt at error mitigation or improved sampling   \cite{Ying-2017, Kandala-2019, Sugura-2018, Arute-2020}. Compared to the exact/simulation results (solid lines), the moments computed on the quantum computer system are surprisingly free of shot noise, with deviations largely due to the device errors. The cumulants have higher statistical noise, as expected given their composition in terms of the moments. In Fig \ref{Fig4}(d) we plot the infimum estimates $E_0^{(\rm inf)}$ obtained from the device runs together with variational results on $\langle H\rangle_{\theta}$ and simulations carried out for different noise levels (zero to 8$\times$ device default error model). We make the following observations: 

\noindent (i) The infimum estimate significantly improves on the variational result for the same trial-state; 

\noindent(ii) Despite the classical manipulation of measured quantities to assemble $\langle H^n\rangle$ and $c_n$, the overall statistical noise in the final QCM infimum results appears to be not too much greater than the variational results, and certainly much less than their difference; 

\noindent(iii) The quality of the infimum estimate derived from the trial-state on the 1D qubit array persists for a range of values of the trial-state parameter either side of the variational minimum ($\theta=\pi$);

\noindent(iv) The simulations indicate the infimum estimate is more robust to device noise than the variational calculation on $\langle H\rangle_{\theta}$. 

To test these observations we move on to larger and more complex instances of the 2D model. The 1D trial-state form $|\phi_{\rm trial}(\theta)\rangle$ is retained, but the model is generalised to the case of random couplings, $\{J_{ij}^{(x,y,z)}\}$. We note another important feature of the QCM approach -- once the Pauli string reduction and measurements have been carried out for a particular problem graph, one need not repeat when the couplings in the problem Hamiltonian are changed. In effect, one only needs to run the moments computation once on the quantum computer -- the infimum estimates for an arbitrary large ensemble of random instances can be computed efficiently using classical resources {\it post-facto} by recycling the quantum computed moments output. For problem instances up to $4\times 4$ we compute and compare with exact results, however, at $5\times 5$ the Hilbert space dimension of the problem is $O(10^7)$ and begins to challenge convenient classical computation. As a reference, we compare with the 2D Heisenberg model case with uniform couplings for which the ground-state is known numerically   \cite{Xiang-2001} (Note the infinite lattice limit value is $E_0 = -2.676$ in our qubit notation). 

\begin{figure*}[htb]
\includegraphics[width=17cm]{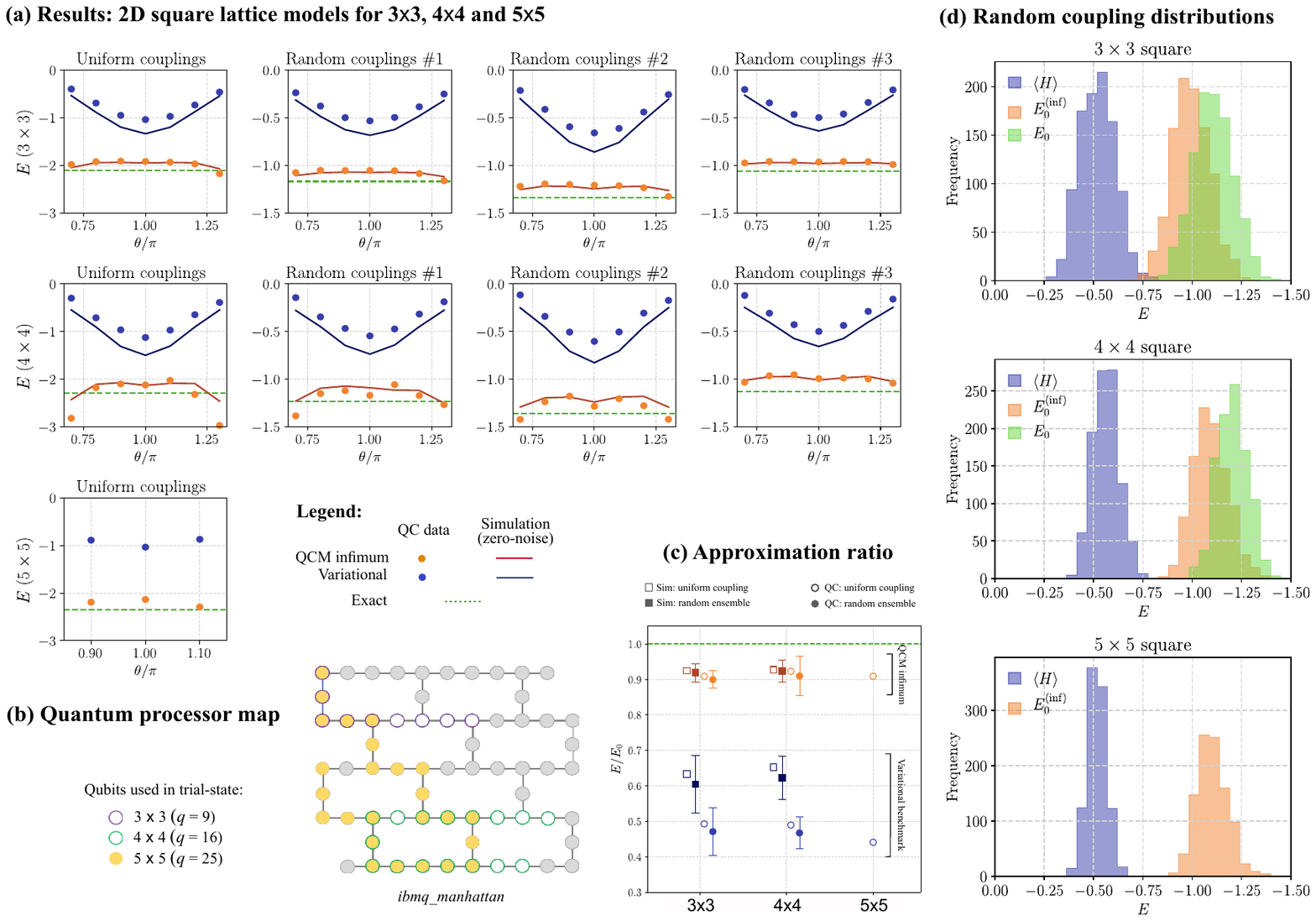}
\centering
\caption{Comparison of QCM infimum and variational benchmark estimates for the generalised 2D Heisenberg model on square lattices of increasing size. (a) Experimental results obtained from the IBM Quantum devices {\it ibmq\_manhattan} and {\it ibmq\_toronto} ({\it ibmq\_toronto} data for $4\times 4$: $\theta=0.7\pi$ and $\theta=1.3\pi$), with 8192 shots per point, plotted as a function of $\theta$. Solid lines correspond to zero-noise simulations (Qiskit QASM simulator) for variational (blue) and infimum (orange) estimates. The exact ground-state energy is plotted as a green dashed line. For each lattice size the trial state $|\phi_{\rm trial}(\theta)\rangle$ (Fig \ref{Fig3}(b)) was encoded using a linear chain of qubits, as shown in the device map (b). (c) Approximation ratios with respect to the exact results for the uniform coupling model and the random coupling ensemble average ($10^3$ instances). (d) Frequency distribution over the random coupling ensemble for QCM infimum estimates and variational benchmark, compared to the exact results calculated for $3\times 3$ and $4\times 4$.}
\label{Fig5}
\end{figure*}

In Fig \ref{Fig5}(a) we show the results for square lattices $3\times3$, $4\times4$, and $5\times5$ (8192 shots per data point) with the trial-state implemented on qubit chains as shown on the IBM Quantum device {\it ibmq\_manhattan} Fig \ref{Fig5}(b). Random coupling instances correspond to choosing $J_{ij}^{(x,y,z)}\in[0,1)$ (three decimal places). Across the board, the QCM infimum results improve on the variational benchmark and are remarkably close to the exact results given the 1D restriction of $|\phi_{\rm trial}(\theta)\rangle$. The data is accompanied by zero-noise simulations, which clearly show that while the variational data points obtained with device noise consistently move away from the true ground-state energy, the QCM infimum estimates around the N\'eel point are remarkably inert and maintain the robustness to both noise and change in the trial-state as shown in the $q = 6$ results. In Fig \ref{Fig5}(c) we plot the approximation ratio with respect to the exact result. The difference between the QCM results and the variational benchmark is clear -- device and shot noise are well under control for the QCM infimum estimates. For the largest $5\times5$ instance, the approximation ratio with respect to the exact value (uniform 2D Heisenberg model \cite{Xiang-2001}) for the QCM infimum estimates is 91\%. The recycling of the one-time quantum output also works well, despite the assembly process for $\langle H^n\rangle$ and $c_n$, providing an ensemble of results over random coupling instances -- the ensemble distributions ($10^3$ instances) are shown in Fig \ref{Fig5}(d).

\section{Conclusion} \label{sec:Conclusion}
The Quantum Computed Moments approach presented here shifts the focus of the representation of problem complexity from the trial-state to the quantities being measured on the quantum computer -- Hamiltonian moments. We demonstrated the method on models of 2D quantum magnetism using IBM Quantum processors for instances up to 25 qubits. At order $\langle H^4\rangle$, the data suggests the classical pre-processing into TPB scales sub-linearly for these models. For our investigations a single-parameter trial-state was chosen, implemented over relatively shallow depth circuits on the devices at hand, and which encompassed the antiferromagnetic N\'eel state. Hamiltonian moments to fourth order were quantum computed and the infimum estimate obtained was found to provide a significant correction to the variational estimate. The infimum results were stable over a significant range of the variational parameter either side of the minimum $\langle H\rangle$ (N\'eel point). The $5\times 5$ instances begin to surpass convenient classical verification, however, our results compared well to those reported in the literature for the uniform coupling case \cite{Xiang-2001}. Our trial-state was chosen as a simple benchmark for comparison purposes, rather than the quest for precision in the final result. Given the relative robustness of the infimum results to device noise we expect there is scope to improve the precision further through more carefully designed trial-states and/or error mitigation on the measured quantities \cite{Ying-2017, Kandala-2019}. On the question of relative computation workload on the quantum processor: for the benchmark trial-state used here the classical computation of the moments can be cast as an efficient sampling problem given the structure of the trial-state circuit \cite{IQP-2011}. More complex trial-states can be constructed, however, our focus has been on demonstrating the robustness of the approach -- in particular that away from the N\'eel point at $\theta=\pi$ where the trial-state is more populated, the estimates clearly survive the effect of device errors and shot noise on the arithmetical processes. Finally, we demonstrate an important feature of the hybrid calculation -- once the TPB measurements are carried out on the quantum computer, the infimum estimate for any other Hamiltonian of the same form can be computed entirely in the classical post-processing. Our results for random coupling instances are obtained by recycling a one-time set of quantum measurements over the single-parameter trial state. Even though the trial-state is more suited to uniform couplings, where we were able to perform exact calculations, the relative precision of the QCM infimum results consistently holds as per the uniform Heisenberg model case. 

In this work we have focused on the practicalities of the quantum computation of Hamiltonian moments to produce estimates of the ground-state energy of a given problem. In introducing the moments based approach we have demonstrated the relative improvement possible over variational calculations for the same relatively simple trial-state. In principle, higher order approximates can be obtained from the cumulant expansion. Further studies would include a systematic analysis of different trial-states on the precision of the estimates obtained and the potential to provide stable results within the quantum volume \cite{Cross-2019} constraints of state-of-the-art devices. Of particular interest, relevant to the question of quantum  advantage, would be the question of simulatability of the moment calculation with respect to the quantum volume and degree of entanglement in the trial-state. In terms of scaling, we have shown in the lattice spin problem considered that the TPB set reduction offers significant savings in both classical computing requirements in the pre-processing step, and the number of measurements on the quantum system. For problems with a large number of terms in the base-level Hamiltonian, e.g. in quantum chemistry systems which scale as $N^4$ with respect to the number of orbitals $N$, we expect the naive term count in $H^4$ to similarly drop significantly in the reduction process, however, further work is required to determine the degree of compression in such cases. Additionally, perturbative approaches and/or cut-off thresholds could be usefully applied to maintain a workable scaling in $H^n$. Expectation values of other quantities of interest can be obtained in a Hellmann-Feynman approach by computing the ground-state energy in the presence of the appropriate term in the Hamiltonian \cite{LH-1994a}. Going beyond ground state energy problems, there is scope for determining the solution configuration through the Lanczos approach, as well as the application of similar moment based results for excited states \cite{LH-1995}, and the application to $ZZ$ optimisation problems by systematic inclusion of mixing terms in the corresponding Hamiltonian form of the problem -- thereby expanding the class of problems the approach could be applied to. 

\section{Contributions and acknowledgements} \label{sec:ack}
The quantum computed moments approach was conceived by LCLH. Early investigations of the method were carried out by MAJ in a MSc project (University of Melbourne, 2019), under the supervision of LCLH and CDH. HJV and LCLH conducted the computations and analysis in this work. LCLH wrote the paper with input from the authors. The authors thank S. Tonetto for helpful discussions and F. Creevey for assistance with HPC aspects. The research was supported by the University of Melbourne through the establishment of the IBM Q Network Hub at the University.

\end{document}